\documentclass[11pt,twoside]{article}
\usepackage{asp2014}
\usepackage{ulem}
\usepackage{xcolor}

\aspSuppressVolSlug
\resetcounters

\begin{document}

\title{FAIR solutions for a science platform to analyse Cherenkov data online}

\author{Mathieu~Servillat,$^1$ Paula~Kornecki,$^1$ Catherine~Boisson$^1$}
\affil{$^1$LUTH, Observatoire de Paris, Université PSL, Université de Paris, CNRS, F-92190 Meudon, France; \email{mathieu.servillat@obspm.fr}}

\paperauthor{Mathieu~Servillat}{mathieu.servillat@obspm.fr}{0000-0001-5443-4128}{LUTH - Observatoire de Paris, CNRS}{}{Meudon}{}{92190}{France}
\paperauthor{Paula Kornecki}{paula.kornecki@obspm.fr}{0000-0002-2706-7438}{LUTH - Observatoire de Paris, CNRS}{}{Meudon}{}{92190}{France}
\paperauthor{Catherine Boisson}{catherine.boisson@obspm.fr}{0000-0001-5893-1797}{LUTH - Observatoire de Paris}{}{Meudon}{}{92190}{France}



  
\begin{abstract}

We developed a system to run quick analyses of Cherenkov data in compliance with the FAIR Guiding Principles for scientific data management (FAIR: Findable, Accessible, Interoperable and Reusable), through the use of interoperability standards and technologies, particularly those provided by the International Virtual Observatory Alliance (IVOA) to build the Virtual Observatory (VO).

We therefore provide a controlled and stable environment on a computing cluster, in order to execute and re-execute well defined jobs. User-specific input parameters can be specified to configure the execution of an analysis job. Provenance information is automatically captured by the system and accessible to the user. To avoid long transfers, the data can be placed close to the computing nodes. This system is primarily used to analyse Cherenkov astronomy data, though it may be used for other purposes.
  
\end{abstract}





\section{Introduction and context}

In the context of Open Science, data providers include more and more in their requirements the distribution of FAIR data (FAIR: Findable, Accessible, Interoperable and Reusable), as described by the FAIR Guiding Principles for scientific data management \citep{Wilkinson2016}. However, transforming the FAIR Principles into working implementations of a data access system is not always straightforward. 

We propose an implementation where data access, job definitions, and computing resources are separated modules, and use standards to implement the interfaces. Data access is based on the IVOA Table Access Protocole (TAP, \citealt{2019ivoa.spec.0927D}), job definitions are compatible with the IVOA Provenance Data Model \citep{2020ivoa.spec.0411S}, and the execution of jobs on computing resources follow the IVOA Universal Worker Service pattern (UWS, \citealt{2016ivoa.spec.1024H}). Authentication is possible through the OpenID Connect standard, and user management follows the SCIM standard (System for Cross-domain Identity Management).

\section{From principles to implementation}
\label{sec:principles}

\subsection{Findable}

\begin{itemize}
\setlength{\parskip}{2pt}
\setlength{\itemsep}{2pt}

    \item Data should be described with \textit{standard metadata}: we provide metadata based on the IVOA Observation Core Metadata model \citep{2017ivoa.spec.0509L} adapted to Cherenkov event data, and suitable for high energy astronomy.

    \item Metadata is served with a \textit{standard protocol}: we use an IVOA TAP server based on GAVO DaCHS \citep{2014A&C.....7...27D} that answers to ADQL queries.

    \item This metadata service is \textit{publicly declared} in the VO Registry \citep{2018ivoa.spec.0723D}.

\end{itemize}

\subsection{Accessible}

\begin{itemize}
\setlength{\parskip}{2pt}
\setlength{\itemsep}{2pt}

    \item Data can be retrieved directly (through link or service): we indicate in the metadata the file type and its \textit{access URL}.

    \item Authentication may be necessary: we use \textit{federated and token based authentication} (OpenID Connect\footnote{\url{https://openid.net/connect}}) to connect to services like KeyCloak (at Observatoire de Paris), eduTeams or ESCAPE IAM.

    \item Data is adequately identified to be found by the computing resources: data is placed in a storage mounted to computing nodes, in order to avoid long file transfers.

\end{itemize}

\subsection{Interoperable}

\begin{itemize}
\setlength{\parskip}{2pt}
\setlength{\itemsep}{2pt}

    \item Interfaces of the data access system are \textit{based on standards}: we use IVOA TAP and UWS as standard interfaces with OPUS \citep{2022ASPC..532..451S}.

    \item Data is stored in a well defined and \textit{standard data format}: a common data format is proposed and discussed for Very-high-energy Observatories (GADF\footnote{\url{https://gamma-astro-data-formats.readthedocs.io}} and now VODF\footnote{\url{https://vodf.readthedocs.io}}).

    \item Related software follow the \textit{FAIR principles for Research Software} \citep{FAIR4RS}: Cherenkov data analyses are performed with the open source software GammaPy \citep{GammapyII2019A&A...625A..10N}.

\end{itemize}

\subsection{Reusable}

\begin{itemize}
\setlength{\parskip}{2pt}
\setlength{\itemsep}{2pt}

    \item Data is presented as complete datasets that contain all relevant pieces of information for further analysis (including instrument response functions).

    \item Data is connected to relevant calibration files and other datasets. We propose to use IVOA DataLink \citep{2015ivoa.spec.0617D} to interconnect data.

    \item Full and \textit{detailed provenance} is available for each dataset: Provenance is automatically captured and exposed following the IVOA Provenance Data Model \citep{2020ivoa.spec.0411S}.

\end{itemize}

\section{Test implementations}
\label{sec:imp}

\begin{figure}[h]
\includegraphics[width=\textwidth]{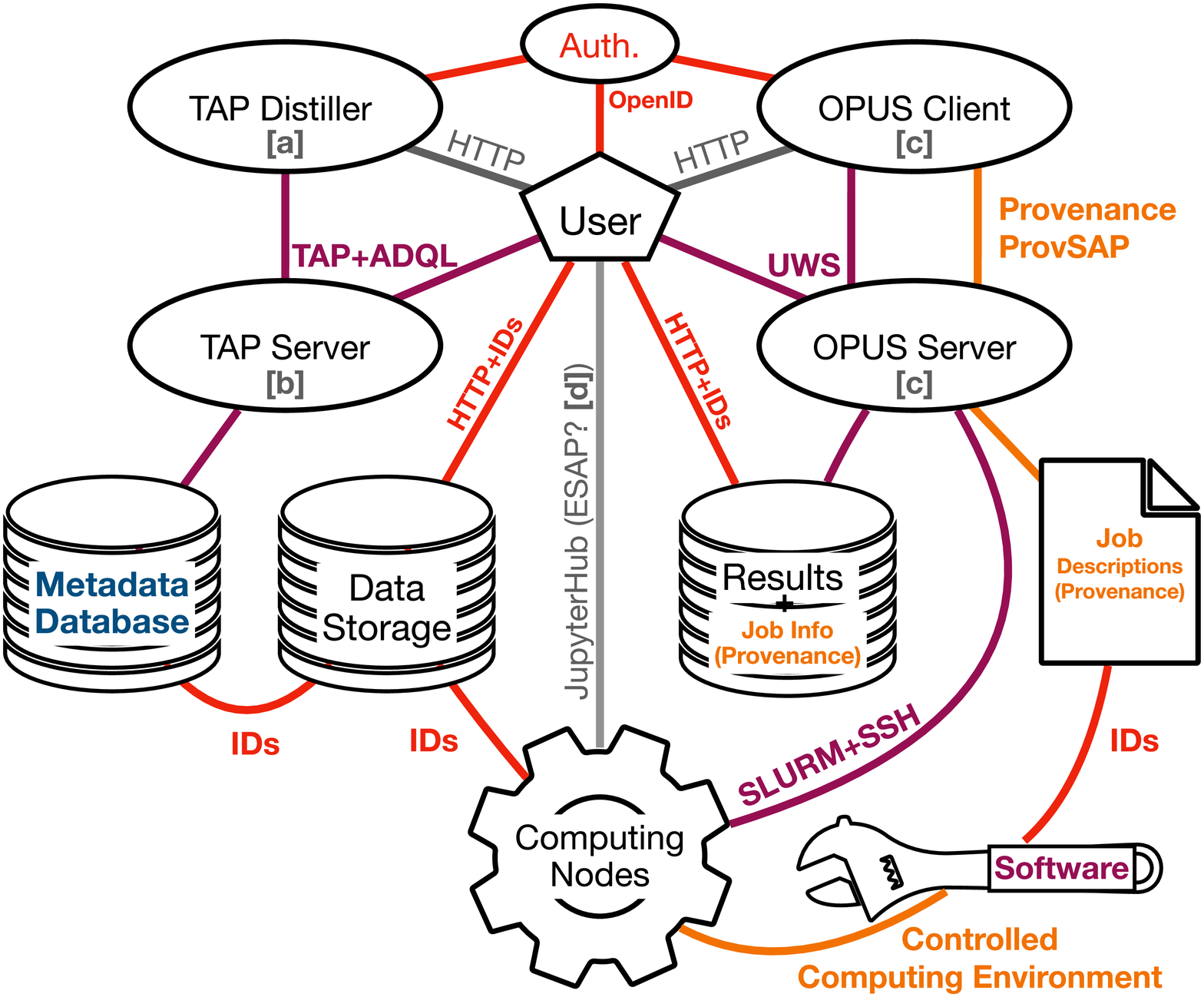}
\caption{Functional diagram of a Science Platform based on the FAIR Principles. Test implementations are further described in \S\ref{sec:imp}. Colors indicate a compliance with the FAIR principles: Findable (blue), Accessible (red), Interoperable (purple) and Reusable (orange).}
\label{fig1}
\end{figure}

Based on technologies and choices listed in Section \S\ref{sec:principles}, we thus developed a data access prototype in line with the FAIR Principles described in Figure~\ref{fig1}, built on the following test implementations:

\begin{itemize}
\setlength{\parskip}{2pt}
\setlength{\itemsep}{2pt}

\item[\textbf{[a]}] \textbf{TAP Distiller}\footnote{\url{https://voparis-cta-test.obspm.fr}}: a web based interface to query a TAP server.

\item[\textbf{[b]}] \textbf{TAP Server}\footnote{\url{http://voparis-tap-he.obspm.fr/browse/hess_dr/q}}: direct access to the TAP server implemented with DaCHS.

\item[\textbf{[c]}] \textbf{OPUS}\footnote{\url{https://voparis-uws-test.obspm.fr}} client/server: UWS interface to run jobs on computing nodes using OPUS \citep{2022ASPC..532..451S}

\item[\textbf{[d]}] \textbf{ESAP}\footnote{\url{https://git.astron.nl/astron-sdc/esap-api-gateway}}: ESFRI Science Analysis Platform developped as a toolkit during the ESCAPE European project.

\end{itemize}

Figure~\ref{fig1} exposes the central position of the user that can authenticate to several web based modules to perform data searches (TAP Distiller) or data analyses online (OPUS CLient). The user can also directly access the TAP server or the OPUS server that implement IVOA standards (TAP+ADQL and UWS). The Data is described by standard metadata distributed by the TAP server. The same data can be accessed by the computing nodes. The OPUS server describes a set of jobs deployed in controlled environments and run on the computing nodes. We currently explore a parallel access to computing nodes and to the data based on a JupyterHub cluster (or a BinderHub cluster), possibly deployed via the ESAP interface.

\section{Working use case}

The FAIR solutions for a science platform were further tested to run a quick gammapy analysis of Cherenkov data online, as presented in \cite{Kornecki:2022}.


\bibliography{P49}


\end{document}